\RequirePackage{amsmath}
\documentclass[12pt]{iopart}
\usepackage[utf8]{inputenc}
\usepackage{amsmath}
\usepackage{amsfonts}
\usepackage{amssymb}
\usepackage{graphicx}
\usepackage{hyperref}
\usepackage{bbold, bm}
\usepackage{siunitx}
\usepackage{cite}

\usepackage{soul}
 
\usepackage{color}
\makeatletter
\renewcommand*{\@textcolor}[3]{%
  \protect\leavevmode
  \begingroup
    \color#1{#2}#3%
  \endgroup
}
\makeatother

\newcommand{\comm}[1]{}
\newcommand{\revision}[1]{\textcolor{black}{{#1}}}
\renewcommand{\vec}[1] {{\bm{#1}}}
\newcommand{\ham}{\mathcal{H}}
\newcommand{\ce}[1] {$\mathrm{#1}$}
\newcommand{\abs}[1] {{\lvert}#1{\rvert}}
\newcommand{\braket}[2]{\langle #1{\rvert} #2 \rangle}
\DeclareMathOperator{\sign}{sign}
\newcommand{\up}{\uparrow}
\newcommand{\dw}{\downarrow}

\begin{document}

\title{Rashba coupling and spin switching through surface states of Dirac semimetals}

\author{Yuriko Baba$^{1, 2, *}$, Francisco Dom\'{\i}nguez-Adame$^{2}$, Gloria Platero$^{3}$ and Rafael A. Molina$^{1, \dagger}$}
\ead{$^*$ yuribaba@ucm.es, $^\dagger$ {rafael.molina@csic.es}}

\address{$^{1}$ Instituto de Estructura de la Materia, IEM-CSIC, Serrano 123, E-28006 Madrid, Spain}
\address{$^{2}$ GISC, Departamento de F\'{\i}sica de Materiales, Universidad Complutense, E-28040 Madrid, Spain}
\address{$^{3}$ Instituto de Ciencia de Materiales de Madrid, ICMM-CSIC, E-28049 Madrid, Spain}
\address{$^{*, \dagger}$ Authors to whom any correspondence should be addressed.}


\vspace{2pc}

\date{\today}

\begin{abstract}

We study the effect of the Rashba spin-orbit coupling on the Fermi arcs of topological Dirac semimetals. The Rashba coupling is induced by breaking the inversion symmetry at the surface. Remarkably, this coupling could be enhanced by the interaction with the substrate and controlled by an external electric field. We study analytically and numerically the rotation of the spin of the surface states as a function of the electron's momentum and the coupling strength. Furthermore, a detailed analysis of the spin-dependent two-terminal conductance is presented \revision{in the clean limit and with the addition of a random distribution of impurities.} Depending on the magnitude of the quadratic terms in the Hamiltonian, the spin-flip conductance may become dominant, thus showing the potential of the system for spintronic applications\revision{, since the  effect is robust even in the presence of disorder.}

\end{abstract}


\maketitle

\section{Introduction}

In crystals that are not invariant under spatial inversion, the energy bands present splitting due to the spin-orbit coupling (SOC). Dresselhaus \cite{Dresselhaus1955} and Rashba \cite{Rashba1960} realized that in noncentrosymmetric semiconductors, bulk SOC becomes odd in momentum. A bit later, Vas'ko \cite{Vasko1979} and Bychkov and Rashba \cite{Bychkov1984} applied this idea to two-dimensional electron gases without inversion symmetry. When the SOC is non-negligible, electrons moving in an electric field experience a magnetic field in their frame of reference that couples to the electron's magnetic moment. This coupling, known as Rashba spin-orbit coupling (RSOC), is essential for many spintronics applications after the pioneering proposal by Datta and Das of a spin field effect transistor \cite{Datta1990}. 
\comm{\revision{\textst{This type of transistor has been experimentally achieved in a semiconductor with strong RSOC connected to ferromagnetic leads and a gate-controlled precession angle of the injected electron spin. The RSOC is particularly attractive for its applications in spintronics and in quantum computation, due to its high tunability.}}}
The strength of the RSOC is directly related to the potential drop at the interface so it can be controlled by an external gate voltage. This capability has already been experimentally demonstrated in InGaAs/InAlAs heterostructures~\cite{Nitta1993} and HgTe quantum wells~\cite{Schultz1996}.

The discovery of the spin Hall effect in topological insulators has opened new avenues of research and possibilities for spintronics applications \cite{Kane2005a,Bernevig2006,Maciejko2011}. The counterpropagating edge states of a bidimensional topological insulator have opposite spin polarization and wave numbers at each boundary, forming Kramers pairs. These states are named helical edge states due to their connection between spin and propagation direction. Time-reversal symmetry prohibits elastic backscattering from one state to its Kramers companion. As these states are the only ones accessible inside the gap, quantized conductance is expected at low enough temperatures. This property was, in fact, observed in the first experiments on topological insulators \cite{Konig2007}. Materials with strong bulk SOC are required for the band inversion needed for the topological insulator state. However, if there is an inversion symmetry breaking, that can be provided by an external field or a substrate, for example, new SOC terms appear (RSOC and bulk inversion asymmetry terms). These terms can break the axial spin symmetry of the edge states. Then, edge states acquire an energy-dependent spin orientation, although they remain Kramers pairs~\cite{Rothe2010,Ostrovsky2012,Rothe2014,Ortiz2016GenericInsulator}. In three-dimensional topological insulators, the splitting of surface states bands by RSOC has been observed experimentally \cite{Zhang2010} and modelled theoretically \cite{Shen2010} for Bi$_2$Se$_3$ on a SiC substrate. 

Gapless three-dimensional topological semimetals have been recently the focus of a lot of attention \cite{Xu2015DiscoveryArcs,Armitage2018WeylSolids}. Bulk SOC is also  of fundamental importance for the stability of these topological semimetals. The surface states of these materials form the so-called Fermi arcs and exist in open compact curves in the Brillouin zone. The effect of a RSOC in the surface of topological semimetals has not been explored yet, except in the two-dimensional ultrathin film limit \cite{Acosta2019}. The intrinsic effect due to the broken inversion symmetry in the surface, although it should exist by symmetry considerations, is probably small and, to the best of our knowledge, has not been observed in experiments. However, we find that a carefully designed sample with a slab of topological semimetal on top of a substrate with heavy atoms should induced a strong and controllable RSOC in the surface that will impact the Fermi arcs. The study of the possibilities for spin dependent transport of this scenario is the subject of this work. The structure of the paper is as follows. In section~\ref{sec:Models} we introduce the minimal model for a Dirac semimetal and the extra terms needed for more realistic models that can describe materials such as Na$_3$Bi, for example. We study analytically the mixing of the chiralities of the surface states due to RSOC in section~\ref{sec:RSOC}. Then, we study numerically the two-terminal conductance in a slab setup in section~\ref{sec:Transp} \revision{and address the problem of the disorder due to a random distribution of impurities in section~\ref{sec:Disorder}}. We finish with some conclusions and perspectives for future work in section~\ref{sec:Concl}.  

\section{Model Hamiltonians} \label{sec:Models}

\subsection{Minimal model for a Dirac semimetal} \label{subsec:MM}

In order to study the effect of the RSOC, we propose two models. The first one, which we will refer to as minimal model, is a four-band Hamiltonian with a pair of Dirac cones in an isotropic and particle-hole symmetric scenario. It had been studied extensively in the literature \cite{Gonzalez2017, Shen2017_Book}, giving a valuable insight in spite of the simplicity of its formulation. 

\begin{figure}[ht]
    \centering
    \includegraphics[width=0.35\textwidth]{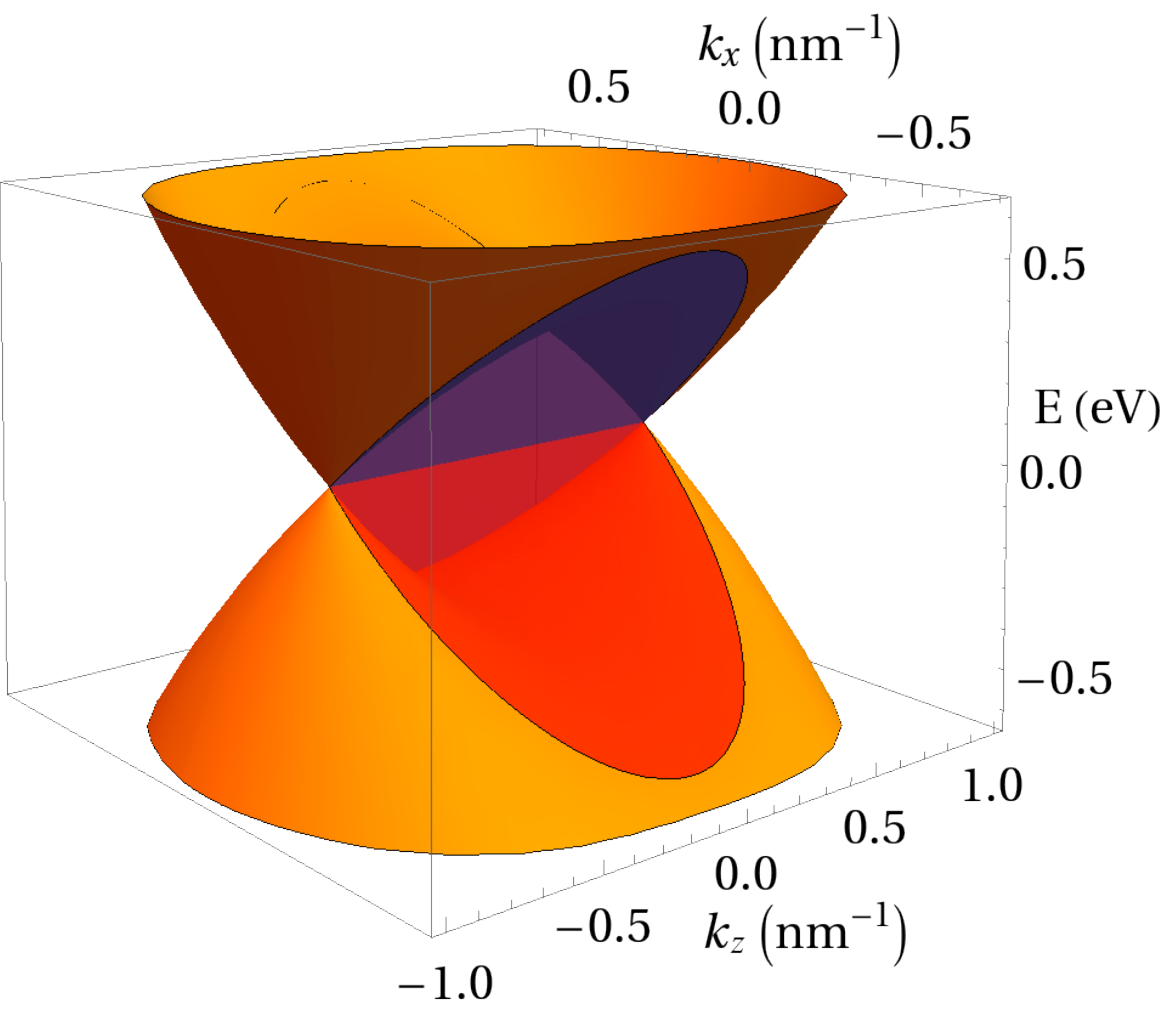}
    \caption{Dispersion relation for $k_y=0$. Bulk states [see equation~\eqref{eq:MM:bEns}] are plotted in orange, surface states [see equation~\eqref{eq:MM:SurfE}] in opaque red and light-blue for $ \zeta = -1$ and $ \zeta = 1$, respectively. The parameters are $m_0=0.35 \si{~eV}$, $m_1=1.0 \si{~eV nm^2}$, and $v= 1.0 \si{~eV nm}$.}
    \label{fig:MM:Disp}
\end{figure}

In the absence of coupling between both chiralities, the Hamiltonian for a Dirac semimetal can be written in two blocks, which are time-reversed partners and are $2 \times 2$ Weyl Hamiltonians with chirality $\zeta$. Therefore, the Hamiltonian has the following form
%
%
\begin{equation} \label{eq:MM:HamD}
    \ham_{D}(\vec{k})=\left( 
    \begin{array}{cc}
    \ham_{W,+1}(\vec{k}) & 0 \\
    0 & \ham_{W,-1}(\vec{k})
    \end{array} \right) ~.
\end{equation}
In the case of the minimal model, each block is obtained from the following Weyl Hamiltonian, \revision{where, in the following, $\hbar =1$ for convenience}
\begin{equation}  \label{eq:MM:HamW}
    \revision{ \ham_{W, \zeta} (\vec{k})=  
    (m_0-m_1 \hbar^2 \vec{k}^2)\sigma_z +\hbar v (\zeta k_x \sigma_x - k_y\sigma_y)~,}
\end{equation}
where $\zeta$ takes on the values $\pm 1$ depending on the chirality and $\sigma_i$ with $i=x,y,z$ are the Pauli matrices. 
The bulk dispersion relation is given by 
\revision{
\begin{equation}  \label{eq:MM:bEns}
    E_b = \pm \sqrt{(m_0 - m_1 \vec{k}^2)^2 + v^2 (k_x^2 + k_y^2)} \ ,
\end{equation}
}%
and it is shown in figure~\ref{fig:MM:Disp}. The valence and conduction bands touch at the Dirac nodes that are located at 
\revision{$\pm\vec{k}_D= \pm (0,0,R)$} with $R=\sqrt{m_0/m_1}\,$.

For the analytic approach, we consider a semi-infinite geometry in the perpendicular direction to the segment joining both nodes, i.e. a semi-infinite slab with a single surface located at $y = 0$ and extended in the half-plane $y>0$. In the other spatial directions, the slab is considered infinite and, therefore, in our analytical calculations, $k_x$ and $k_z$ are good quantum numbers. In order to find surface states that fulfil Dirichlet boundary conditions, we impose $\psi |_{y = 0} =0$. Using the \textit{ansatz} $\psi_{s} \sim e^{i k_x x} e^{i k_z z} e^{-\lambda y} \Phi$, where $\Phi$ is a constant and normalized spinor, we get two solutions
\begin{equation} \label{eq:MM:SurfS}
\psi_{s, \zeta} = A_s \left(e^{-\lambda_1 y}-  e^{-\lambda_2 y}\right) \Phi_\zeta ~, 
\qquad
\zeta=\pm 1\ ,
\end{equation}
where $A_s$ is a normalization constant. $\lambda_1$ and $\lambda_2$ can be real or complex valued and are given by
\begin{equation}
    \lambda_{1}= \Delta + \sqrt{F} \ ,
    \qquad
    \lambda_{2}= \Delta - \sqrt{F} \ ,
    \label{eq:MM:las}
\end{equation}
In the previous expressions we have defined
\begin{equation}
    \Delta\equiv \frac{v}{2m_1} \ ,\qquad F\equiv F(k_x, k_z) = k_x^2+k_z^2-R^2+\Delta ^2~.
    \label{eq:MM:F}
\end{equation}
The spinor in equation~\eqref{eq:MM:SurfS}  is directly related with the chirality of the Weyl semimetal~(WSM) blocks of the Hamiltonian, leading to
\begin{equation} \label{eq:spinors}
    \Phi_{\zeta  = +1 } =  A_\Phi (1, \kappa, 0, 0) ^t, \quad \Phi_{\zeta  = -1 } = A_\Phi (0,0, 1, \kappa) ^t, 
\end{equation}
where $\kappa = 1$, $A_\Phi$ is a normalization constant and the \textit{t} indicates the transpose. These two linearly independent solutions are the celebrated counterpropagating modes, with linear dispersion relation and locked spin
\begin{equation} \label{eq:MM:SurfE}
    E_{s, \zeta} = \zeta v k_x ~.
\end{equation}
More precisely, surface states exists if momenta fulfil the constraint
\begin{equation} \label{eq:MM:CondS}
    {k_x^2+ k_z^2}< R^2~.
\end{equation}
This condition is obtained directly from the definition of the $\lambda_{1,2}$ [see equation \eqref{eq:MM:las}] imposing $\Re{(\lambda_{1,2})}>0$, which implies that $F < \Delta^2$ or, equivalently, we get equation  \eqref{eq:MM:CondS}.

Finally, according to the nature of the decay of the surface states, two phases can be distinguished: The oscillatory decay (i.e. $\lambda_i \in \mathbb{C}$) and the purely exponential decay (i.e. $\lambda_i \in \mathbb{R}$), which we shall denote by type A and type B surface states, respectively. Type A surface states lead to smaller penetration length and oscillatory decay, whereas type B is related to longer and purely exponential decay \cite{Benito-Matias2019,Gonzalez2017}.

\subsection{Na$_3$Bi} \label{subsec:Na3Bi}

In order to elucidate the generality of our results and to study more realistic situations, next we consider the case of a low-energy effective Hamiltonian that describes \ce{A_3Bi} (\ce{A= Na, K, Rb}) \cite{Wang2013Three-dimensionalCd_3As_2} and \ce{Cd_3As_2} \cite{Wang2012DiracRb} around the $\Gamma$ point. These compounds have a single band inversion occurring near the $\Gamma$ point that has been observed by ARPES measurements \cite{Neupane2014ObservationCd_3As_2, Liu2014a, Liu2014, Xu2015}. We use a four-band model that has been obtained as a low energy description of Density Functional Theory (DFT) results \cite{Wang2013Three-dimensionalCd_3As_2, Wang2012DiracRb} for the cited compounds, in which the band inversion provides the formation of a Dirac semimetal (DSM), or more precisely, a $\mathbb{Z}_2$ WSM \cite{Gorbar2015DiracSemimetals, Gorbar2015SurfaceRb}, due to the up-down parity symmetry protection. For the sake of concreteness, we restrict our analysis to the case of \ce{Na_3Bi}.

In the absence of coupling terms between chiralities, the Dirac Hamiltonian has the block-diagonal form of equation~\eqref{eq:MM:HamD} with a WSM Hamiltonian now defined by
\begin{equation} \label{eq:Na3Bi:HamW}
\ham_{W, \zeta} (\vec{k}) = \epsilon_0(\vec{k}) \mathbb{1}_2 + M(\vec{k}) \sigma_z +  v (\zeta k_x \sigma_x- k_y\sigma_y)~,
\end{equation}
where $\epsilon_0(\vec{k})=c_0+c_1 k_z^2+c_2(k_x^2+k_y^2)$, $ M(\vec{k})=m_0-m_1 k_z^2-m_2(k_x^2+k_y^2)$ and $\mathbb{1}_n$ stands for the $n\times n$ unit matrix. The former Hamiltonian has two nodes and terms up to quadratic order in momentum. It generalises the minimal model \eqref{eq:MM:HamW} by allowing for anisotropy in the $z$-direction and the diagonal momentum-dependent term $\epsilon_0(\vec{k})\mathbb{1}_2$. 

Surface states of the model have been discussed in detail in reference~\cite{Benito-Matias2019}. Imposing Dirichlet boundary conditions and a single surface at $y=0$, a surface state has the same form of equation \eqref{eq:MM:SurfS} with the spinor part given by equation \eqref{eq:spinors}, with $\kappa=\sqrt{(m_2-c_2)/(m_2+c_2)}\,$.
Also $\lambda_{1,2}$ have an equivalent form and can be written as in equation~\eqref{eq:MM:las} replacing $F$ and $\Delta$ by the following expressions
\begin{equation}
F_\zeta (k_x, k_z) \equiv \left(k_x +  \zeta k_{x,0}\right)^2 +\frac{m_1}{m_2}\,k_z^2 + \Delta^2- R^2~, 
\quad
\Delta \equiv \frac{v}{2 \sqrt{m_2^2-c_2^2}}~,
\label{eq:Na3Bi:F}
\end{equation}
where $k_{x,0} \equiv {c_2}\Delta/{m_2}$ and $R^2 \equiv {m_0}/{m_2}+ \Delta^2\left({c_2}/{m_2}\right)^2$. Notice that in this model $F_\zeta$ is chiral-dependent. This leads to chiral-dependent regions of existence of surface states that are ellipses in the $k_x-k_z$ plane and have different centers depending on the chirality (see reference~\cite{Benito-Matias2019} for further details).

In contrast to the previous case, the diagonal term $\epsilon_0(\vec{k})\mathbb{1}_4$ leads to a dispersion that is no longer flat along the $z$-direction. Instead, surface states now have the following dispersion
\begin{equation} \label{eq:Na3Bi:SurfE}
E_{s, \zeta} = \varepsilon(k_z)
+\zeta v C_3k_x \ ,
\end{equation}
where the non-flat band contribution is given by
\begin{equation}
    \varepsilon(k_z) = C_1+C_2 k_z^2 \ .
    \label{eq:Na3Bi:dispz}
\end{equation}
In the two last equations $C_1,C_2$ and $C_3$ are a combination of the Hamiltonian parameters. To be specific $C_1 = c_0+c_2 {m_0}/{m_2}$,  $C_2=c_1-c_2 {m_1}/{m_2}$ and $\sqrt{m_2^2-c_2^2}/m_2$. According to the parameters given by reference~\cite{Wang2012DiracRb} and listed in table \ref{tab:Na3Bi}, surface states of the \ce{Na_3Bi} are only of type B. 

\begin{table}[ht]
    \begin{center}
    \begin{tabular}{l l l} 
    \hline
         $c_0 = {-0.06382} \si{~\eV}$  &
         $m_0 = -0.08686 \si{~ eV}$ &
         \\
         $c_1 = 8.7536 \si{~ eV.\angstrom^2}$ &
         $m_1 = -10.6424 \si{~ eV.\angstrom^2}$ &
         $v = 2.4598 \si{~ eV.\angstrom}$
         \\
         $c_2 = -8.4008 \si{~ eV.\angstrom}$ &
         $m_2 = -10.3610 \si{~ eV.\angstrom}$ &
         \\
    \hline
    \end{tabular}
    \end{center}
    \caption{Values for the parameters of \ce{Na_3 Bi}, according to reference~\cite{Wang2012DiracRbb}.}
    \label{tab:Na3Bi}
\end{table}

\section{Impact of the RSOC on the surface states} \label{sec:RSOC}

In this section we study the effect of the RSOC interaction on the previously obtained surface states. The effect of RSOC and spin axial symmetry breaking have been previously studied in two-dimensional~(2D) topological insulators, from the modification of the surface states \cite{Rothe2010FingerprintWells, Ortiz2016GenericInsulator, Schmidt2012InelasticChannel} to the effects on electron  transport \cite{Rothe2010FingerprintWells, Krueckl2011SwitchingConstrictions,Acosta2019}. In a three-dimensional system comprising a slab of  DSM, an equivalent effect is expected due to the interaction between the sample and the substrate. In this scenario the RSOC is a local interaction near the surface and a 2D RSOC term can be used to model the system near the substrate. Considering only the leading linear terms, the Hamiltonian has the following form \cite{Rothe2010FingerprintWells,Acosta2019}
\begin{equation} \label{eq:RSOC:HamRSOC}
    \ham_R=\left( 
    \begin{array}{cccc}
    0 & 0 & -i R_0 k_{-} & 0\\
    0 & 0 & 0 & 0\\
    i R_0 k_{+} & 0 & 0 & 0\\
    0 & 0 & 0 & 0
    \end{array} \right)~,
\end{equation}
where $k_{\pm} \equiv k_z \pm i k_x$. $R_0 $ is a parameter that quantifies the strength of the interaction, which is usually related to an external or internal electric field, depending on the material. A detailed derivation of this type of Hamiltonian for Na$_3$Bi thin films is provided in reference~\cite{Acosta2019}. \revision{In the single layer case, reference \cite{Acosta2019} provides a value of $R_0 \sim 0.654 ~\si{eV \angstrom}$  for an electric field $E = 0.1 ~ \si{eV / \angstrom}$ from DFT calculations.
%
%
In spite of the fact that we are dealing with much bigger samples, we consider that this value is reasonable. However the exact strength of the coupling can be only computed with accurate self-consistency methods by taking into account the details of the substrate as well.}

Notice that $\ham_R$ couples the electron bands but not the hole bands. This particle-hole asymmetry is a very general effect since Rashba terms for electrons depend linearly on $k$ while hole terms are cubic in $k$ \cite{Winkler2000}. However, we will not consider these higher order terms in the present model. The coupling between the two electron bands leads to a coupling between chiralities that breaks the axial spin symmetry. In fact, due to the breaking of the aforementioned symmetry, the chiral surface states (CSSs) turn into states with a more generic and intriguing spin structure than merely having opposite and constant spin orientations independently of energy. Time-reversal symmetry still dictates that the two counterpropagating Kramers partners have orthogonal spinors, but it does not require equal spinors at different energies. These states were called general surface states (GSSs) in contrast to the CSSs obtained in the absence of RSOC~\cite{Ortiz2016GenericInsulator}.

To obtain analytical results we study the interaction in a single surface setup as described in the previous section. Moreover, we assume that the GSS are combinations of the CSS only, neglecting the contribution of the bulk bands. This is a good approximation because the interaction is local and the surface states have a very small overlap with the bulk states. Then, from now on we omit the subindex \textit{s} which denotes the surface states and just label them by their momenta ${\bm k}$ and chirality $\zeta$. In this way we obtain an effective Hamiltonian given as
\begin{equation} \label{eq:RSOC:Heff0}
    \ham_{\mathrm{eff}} =\sum_{{\bm k} \zeta} E_{k \zeta} c^\dagger_{{\bm k} \zeta} c_{{\bm k} \zeta}^{} + \sum_{{\bm k} {\bm k}'} \sum_{\zeta \zeta'} \braket{\psi_{{\bm k} \zeta} |\ham_R}{\psi_{{\bm k}' \zeta'} } c^\dagger_{{\bm k} \zeta} c_{{\bm k}' \zeta'}^{}~,
\end{equation}
where $E_{{\bm k} \zeta}$ is the energy of the CSS given by equation~\eqref{eq:MM:SurfE} and \eqref{eq:Na3Bi:SurfE} and the operator $c^\dagger_{{\bm k} \zeta}$ $(c_{{\bm k} \zeta})$ creates (annihilates) a particle in the CSS $\psi_{k \zeta}$, which we will refer in the following as $\psi_{{\bm k}\up}$ and $\psi_{{\bm k}\dw}$, where the up and down arrows correspond to $\zeta = 1$ and $\zeta = -1$, respectively. More explicitly, the effective Hamiltonian reads
\begin{equation} \label{eq:RSOC:Heff}
    \ham_{\mathrm{eff}}=\sum_{\bm k}  
    \left(c^\dagger_{{\bm k} \uparrow}, c^\dagger_{{\bm k} \downarrow} \right)
        \left( \begin{array}{c c}
        E_{{\bm k} \up}
        & - i R_0 g_{\vec{k}} k_{-} \\
        i R_0 g_{\vec{k}} k_{+}  & E_{{\bm k}\dw}
        \end{array} \right)
         \left(\begin{array}{c}
              c_{{\bm k} \uparrow}\\ c_{{\bm k} \downarrow}
         \end{array} \right) ~,
\end{equation} 
where $E_{{\bm k}\up(\dw)}$ is the energy of the CSS in the absence of the RSOC and 
\revision{the off-diagonal part has been written as a function of a momentum-dependent function, $g_{\vec{k}}$, for convenience. The latter is simply obtained from the second term of equation \eqref{eq:RSOC:Heff0} as $g_{\vec{k}} \equiv i \braket{\psi_{{\bm k}\up}|\ham_R}{\psi_{{\bm k}\dw}}/R_0 k_{-}$. Performing the integration and using the states reported in equation \eqref{eq:MM:SurfS}, we get 
}
\begin{equation}
g_{\vec{k}} = 
\frac{16 \abs{A_\Phi}^2 \Delta^2  \sqrt{\Delta^2-F_\dw} \sqrt{\Delta^2-F_\up}}
{16 \Delta ^4-8 \Delta ^2 (F_\dw+F_\up)+(F_\dw-F_\up)^2}~,
\end{equation}
where $\Delta$ and $F_{\up,\dw}$, that corresponds to $F_{\zeta = \pm1}$, have been defined previously for both models. By diagonalizing $\ham_{\mathrm{eff}}$ we get the dispersion relation
\begin{equation}  \label{eq:E_RSOC}
E^{\mathrm{RSOC}}_{\pm}({\bm k})= \varepsilon(k_z) \pm v_{\bm k} \abs{C_3} k_x~,
\end{equation}
with a modified velocity 
\begin{equation}
v_{\bm k} \equiv \sqrt{v^2 +\frac{g_{\vec{k}}^2 {R_0}^2 \left(
 {k_x^2}+{k_z^2}\right)}{{C_3}^2 {k_x}^2}}~,
\end{equation}
where $C_3$ depends on the model parameters, $\varepsilon(k_z)$  is the non-flat band contribution, being zero for the minimal model. For the range of positive $k_x$, the eigenstates, or equivalently the GSS, can be written as
\begin{subequations}
\begin{align} 
    \psi_{{\bm k}\pm} = \frac{1}{\sqrt{2 v_{\bm k} ( v + v_{\bm k})}} 
    \begin{pmatrix}
    v + v_{\bm k}\\ \label{eq:psi_up}
    i g_{\vec{k}} R_0 k_{+}/{C_3 k_x}
    \end{pmatrix}
    ~,\\
    \psi_{{\bm k}\mp} = \frac{1}{\sqrt{2 v_{\bm k}( v + v_{\bm k})}} 
    \begin{pmatrix}
    i g_{\vec{k}} R_0 k_{-}/{C_3 k_x} \\ v  + v_{\bm k}
    \end{pmatrix} ~, \label{eq:psi_dw}
\end{align}
\label{eq:psi_up_dw}%
\end{subequations}
where the upper sign refers to the case of $C_3>0$ and the lower one to models with $C_3<0$. Notice that for $R_0=0$ and $v_{\bm k}= v$ we recover the expected correspondence between GSS and CSS, namely $\psi_{{\bm k}+} = \psi_{{\bm k}\up}$ and $\psi_{{\bm k}-} = \psi_{{\bm k}\dw}$.

The connection between the CSS and the GSS becomes clearer if we write down the effective Hamiltonian \eqref{eq:RSOC:Heff} in the following form
\begin{equation}
\ham_\mathrm{eff}  = \sum_\vec{k} C^\dagger_k \left( h_0 \mathbb{1}_2 + \vec{h}\cdot \vec{\sigma} \right) C_k~,
\end{equation}
where $C_k^\dagger = \left(c^\dagger_{{\bm k} \uparrow}, c^\dagger_{{\bm k} \downarrow} \right)$, $h_0 = \varepsilon(k_z)$ and the Pauli vector $\vec{h}$ is given by
\begin{align}
    \vec{h} = - \abs{C_3} k_x v_k \left( \sin\theta \cos\phi, \sin\theta \sin\phi, \cos\theta \right)~.
\end{align}
The angles are momentum-dependent
\begin{equation}
    \theta = \cos^{-1}\left[-\sign(C_3) v/v_{\bm k}\right]~ , \qquad
    \phi = \tan^{-1}(-k_z/k_x)~.
\end{equation}
Within this formulation, the eigenvectors given by equation~\eqref{eq:psi_up_dw} can be written as 
\begin{equation} 
    \psi_{{\bm k}\pm} = 
    \begin{pmatrix}
    \sin(\theta/2)\\
    -e^{i\phi}\cos(\theta/2)
    \end{pmatrix}
    ~,
    \qquad
    \psi_{{\bm k}\mp} = 
    \begin{pmatrix}
    e^{-i\phi}\cos(\theta/2)\\
    \sin(\theta/2)
    \end{pmatrix} ~.
\end{equation}
%
In the former expression it is apparent that the already mentioned property of the GSSs, namely the pair of surface states still have ortogonal spinors but their spin structure is more complex. In fact it can be described by the two angles $\theta$ and $\phi$, which are momentum dependent. 

\subsection{Minimal model}

In the next two subsections we particularize these results for the minimal model and for \ce{Na_3Bi}. The impact of RSOC on transport phenomena will be discussed later (see section~\ref{sec:Transp}). We start by analyzing the minimal model described in subsection \ref{subsec:MM}. In this model, $\varepsilon = 0$, $C_3=1$, $\kappa = 1$ and $F$ does not depend on chirality. Therefore, $g_{\vec{k}}= 1/2$ and the modified velocity as well as the GSSs have an utterly simplified form. In fact, due the absence of momentum dependence in $g_{\vec{k}}$, the GSSs are simply obtained by rotating the CSSs and, with a redefinition of the spin bases, the states are still well defined chiral states. This can be easily seen by writing down the dispersion of the states of the effective model. In the minimal model, equation~\eqref{eq:E_RSOC} reduces to an anisotropic version of a 2D Dirac equation with an $R_0$-dependent velocity
\begin{equation}
    E^{\mathrm{RSOC}}_\pm({\bm k}) = \pm \sqrt{\left(v^2+\frac{R_0^2}{4} \right) k_x^2+ \frac{R_0^2}{4} k_z^2}~,
\end{equation}

\subsection{Na$_3$Bi}

The case of \ce{Na_3Bi} has some more subtleties due to the non-flat bands. 
In this model, $C_3 = \sqrt{m_2^2-c_2^2}/m_2$ and $g_{\vec{k}}$ is momentum-dependent.
Explicitly, it is given by
\begin{equation}
g_{\vec{k}} =
-\frac{\sign(m_2) c_{2+} \sqrt{{{c_2^2 k_x^2 v^2}/c_{2\pm}^2+\left(k_x^2 m_2+k_z^2 m_1-m_0\right)^2}}}
{2 \left(k_z^2 m_1 m_2-c_{2\pm}^2 k_x^2 - m_0 m_2\right)}~ ,
\end{equation}
where $c_{2\pm} = c_2\pm m_2$ and $c_{2\pm}^2 = c_2^2- m_2^2$. For small momenta, it can be approximated by a parabolic function
\begin{equation}
g_\vec{k} = g_0 + g_1 k_x^2+g_2 k_z^2 + \mathcal{O}(k_x^2, k_z^2)~,
\end{equation}
with $g_{0,1,2}$ constants given by the model parameters. In particular $g_0 = {c_+}/{2m_2}$ resembles the minimal model momentum independent contribution.

\begin{figure}[htb]
    \centering
    \includegraphics[width=0.45\textwidth]{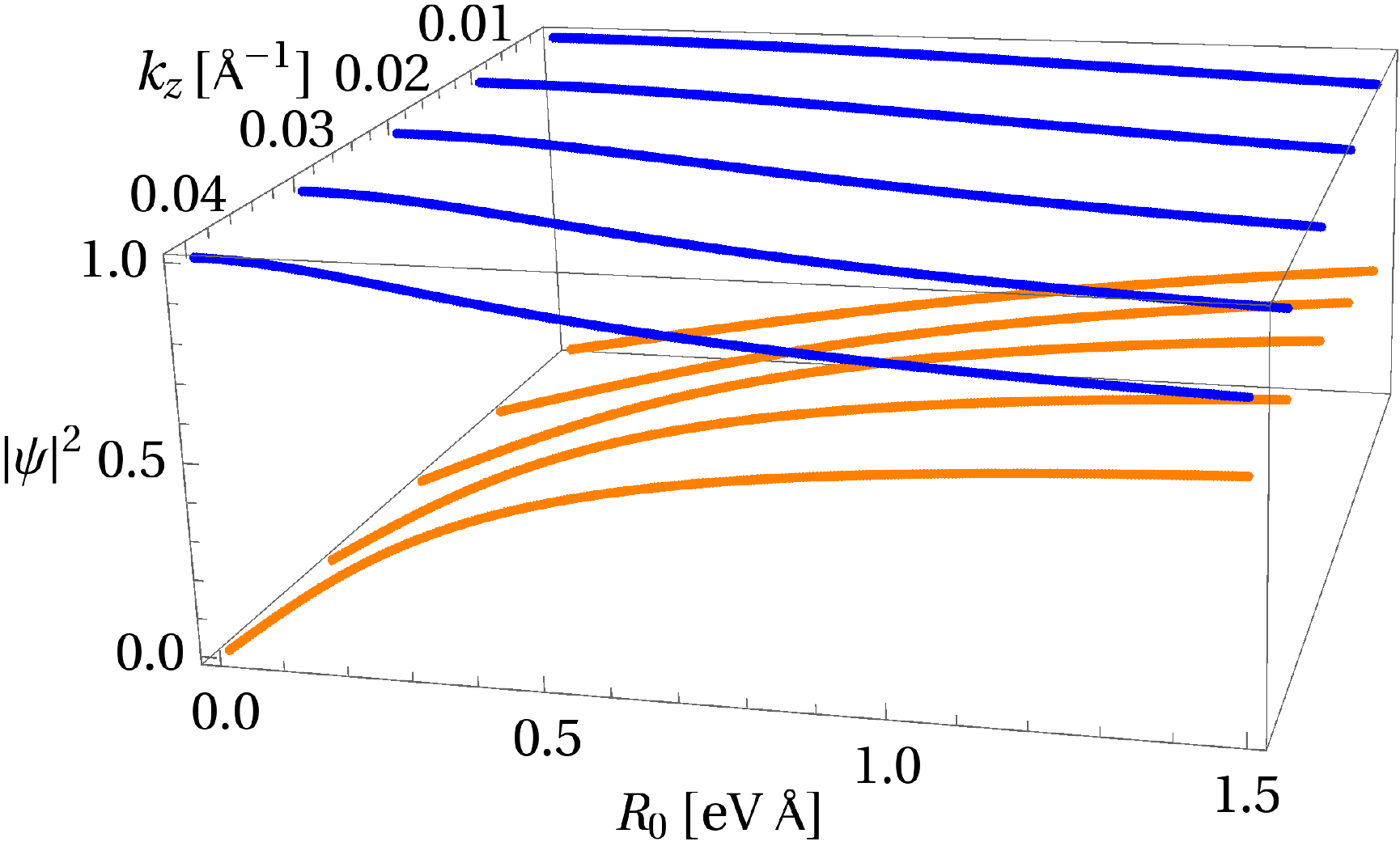}
    \caption{Squared absolute value of the upper (blue) and lower (orange) component of $\psi_{{\bm k}+}$ for \ce{Na_3Bi} as a function of $R_0$ and $k_z$ at fixed $k_x= 0.01 \si{\angstrom^{-1}}$.}
    \label{fig:rNa3Bi:GSS}
\end{figure}

Hence, due to the quadratic terms, the GSSs are more complex and a non-trivial spin-mixing behavior is obtained. Figure \ref{fig:rNa3Bi:GSS} shows the absolute value of the upper and lower component of $\psi_{{\bm k}+}$ as a function of $R_0$, or equivalently as a function of the in-plane electric field. The mixing of the chiralities generates spin-rotation effects that have an important impact on the conductance of the surface states. 

\section{Effect of the RSOC on electron transport in a finite slab setup} \label{sec:Transp}

\begin{figure}[ht]
    \centering
    \includegraphics[height=4 cm]{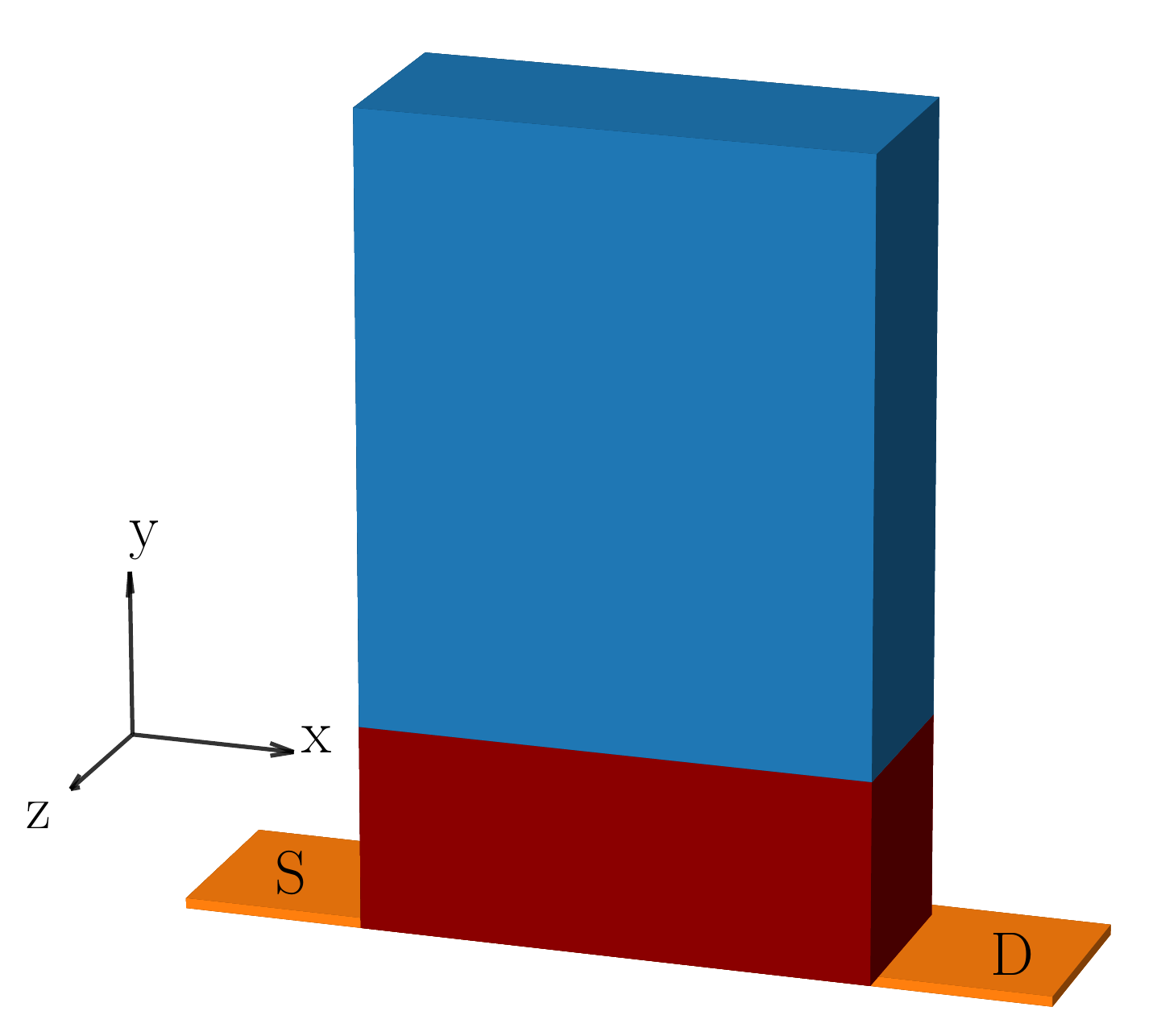}
    \caption{Schematic view of the device. The metallic leads are represented in orange with S and D indicating the source and the drain, respectively. The red color marks the part where the RSOC is considered whereas the blue part indicates the pure DSM without coupling.}
    \label{fig:trans:set-up}
\end{figure}

The mixing of the chiralities opens the possibility of scattering and thereby deviations from quantized spin-conductance of the ideal DSM. Moreover, the spin switch takes place due to the RSOC and interesting effects can occur due to the non trivial spin texture of the scattering states leading to non zero spin-switch conductance in the case of non-trivial spin rotation. To study the effect of the RSOC interaction on the transport, we perform transport simulations applying the Landauer-Büttiker formalism \cite{Datta1997ElectronicSystems} at zero-temperature with the toolkit Kwant \cite{Groth2014Kwant:Transport}. We propose a finite slab system, with 2D metallic leads, which is represented schematically in figure~\ref{fig:trans:set-up}. The choice of this setup aims at measuring  transport properties of the surface states only. A similar setup has been studied in reference \cite{ChestaLopez2018MultiterminalSemimetal} for a WSM. Here, we exploit it to study the spin-polarized electric conductance of the surface states in a DSM with two spin channels.
\revision{The scattering problem in Kwant is solved in an infinite system consisting of the finite scattering cuboid of DSM connected to semi-infinite, in this case bidimensional, leads. The scattering problem is solved using the wave function formulation of the scattering problem implemented in the package, that is equivalent but more efficient than the non-equilibrium Green's function method~\cite{Groth2014Kwant:Transport}. The leads are treated as infinite and they act as wave guides leading plane waves into and out of the scattering region. The main output of the calculation is the scattering matrix of the system $\mathcal{S}_{nm}$ from the $n$ incoming mode to the $m$ outgoing mode.}

\revision{
 The results of the differential conductance presented here are obtained from the Landauer-Büttiker formula particularized to the two-lead device. From the scattering matrix $\mathcal{S}_{nm}$, it reads
\begin{equation}
    G_{i,j} = \frac{e^2}{h}\sum_{n \in i,~m \in j} \abs{\mathcal{S}_{nm}}^2~,
\end{equation}
where the index $i (j)$ labels the modes in the first (second) lead that are eigenstates of a certain operator. 
More precisely, we are mainly concerned with the chiral operator that possesses two eigenvalues with opposite sign. Therefore, labeling the modes by their chirality, we get $G_{\pm \pm}$ and $G_{\pm \mp}$ that corresponds to the conductance of modes with the same chirality and the conductance of modes with opposite chirality in the incoming and outgoing channels, respectively. The total conductance $G_\mathrm{tot}$ is obtained from the polarised ones by
\begin{equation}
    G_\mathrm{tot}= G_{++}+G_{--}+G_{+-}+G_{-+}~.
\end{equation}
The system has been discretised with a lattice spacing of $5 ~\si{\angstrom}$. We have checked that the results are then accurate enough for our purposes.}

\begin{figure}[ht]
    \centering
    \includegraphics[height=5cm]{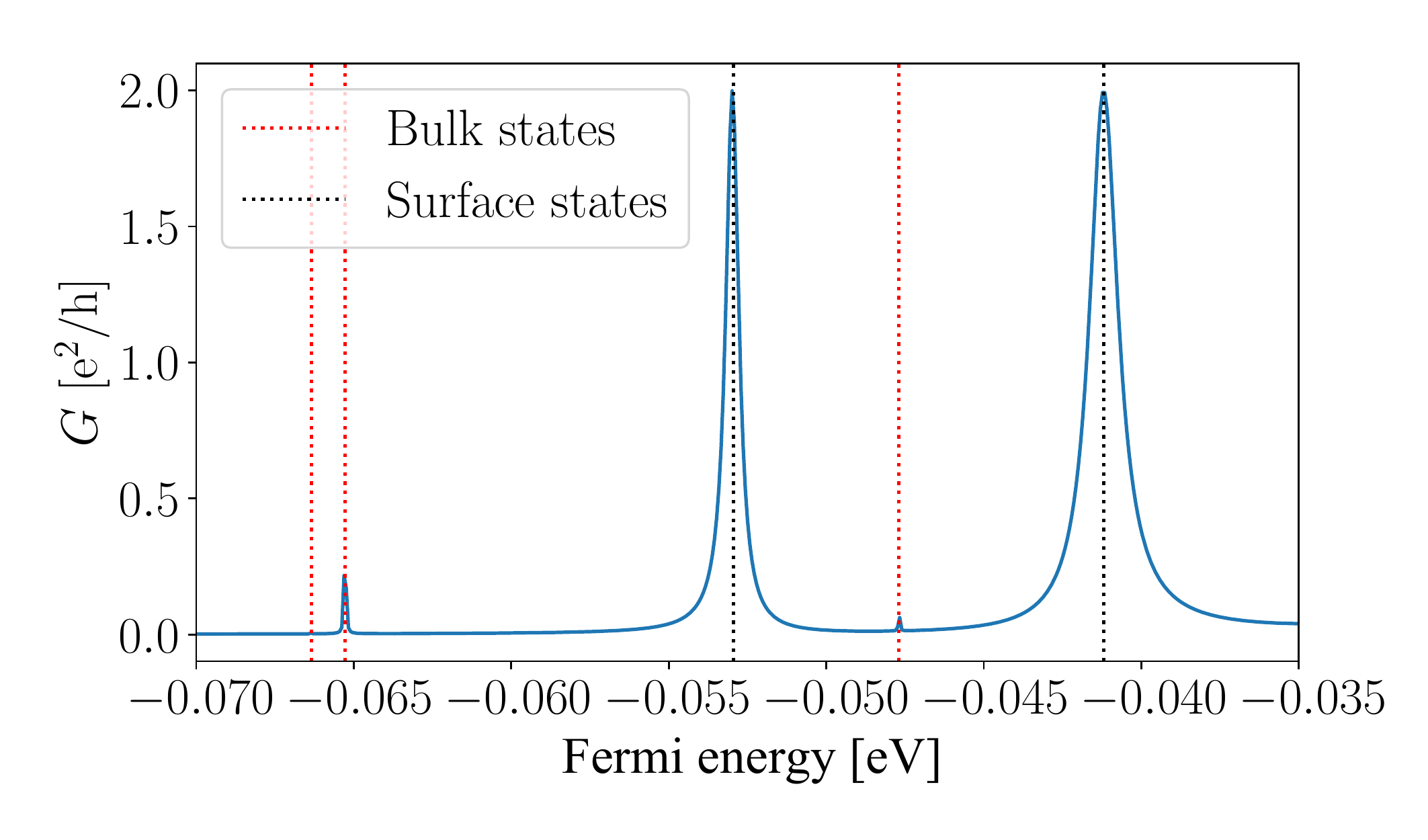}
    \caption{Conductance as a function of the Fermi energy; the sample in the case of \ce{Na_3Bi} for $R_0=0$ and a slab of size ${100\times 250 \times 40}~\si{\angstrom^3}$. The eigenvalues of the system are marked with black (red) vertical lines for surface (bulk) states. }
    \label{fig:trans:bulkVsSurf}
\end{figure}

The details of the setup are the following. The slab has dimensions $L_x \times L_y \times L_z$ with 2D metallic leads attached in the plane $y=0$ at the two ends of the sample. The leads are metallic and intersect the sample at the edge extended along $z$ and are semi-infinite in the $x$ direction. Due to the 2D leads, transport only takes part due to the surface states. Moreover, due to the coupling of metallic leads to a different material, an accurate choice of the hopping parameters of the leads is needed. The leads have a conventional Hamiltonian $\ham_{\mathrm{leads}}= (A\vec{k}^2 + B) \mathbb{1}_4$, where $A$ and $B$ have been chosen to connect efficiently with the sample and to obtain the quantized conductance in the absence of RSOC. Figure~\ref{fig:trans:bulkVsSurf} shows the conductance as a function of the Fermi energy for a slab in the absence of RSOC. The system reaches the expected quantized conductance only for the surface states while the transport is practically absent for the bulk states.

Due to the finite size, the surface states are present at discrete energies and the quantization of $k_x$, $k_y$ and $k_z$ leads to multiple surface states. We set the Fermi energy of the sample to the fundamental mode in the absence of RSOC and we study how the conductance $G$ evolves with $R_0$. This setup is quite close to an experimental realization: The goal is to set the Fermi energy of the sample to the maximum value of transmission for $R_0=0$, which corresponds to the fundamental mode. Then, the RSOC term is applied turning on an electric field in the substrate.  Therefore, we restrict our analysis to this mode, which is nodeless in all spatial directions. We choose this state for the sake of concreteness but we observe similar results in the other surface state modes. In \ce{Na_3Bi} hybrid bulk-surface states arise at higher energies due to the quadratic terms in the Hamiltonian. Therefore, the conductance is not perfectly quantized even in the absence of RSOC. These states are strongly size-dependent but their contribution can be avoided by choosing properly the Fermi energy, as shown in figure~\ref{fig:trans:bulkVsSurf}.

\subsection{Spin-switcher}

In the following subsections we particularize the results for the minimal model and for \ce{Na_3Bi}. As expected, the non-trivial mixing in \ce{Na_3Bi} have very different consequences in the transport properties compared to the minimal model.

\begin{figure}[htb]
    \centering
    \includegraphics[height=5cm]{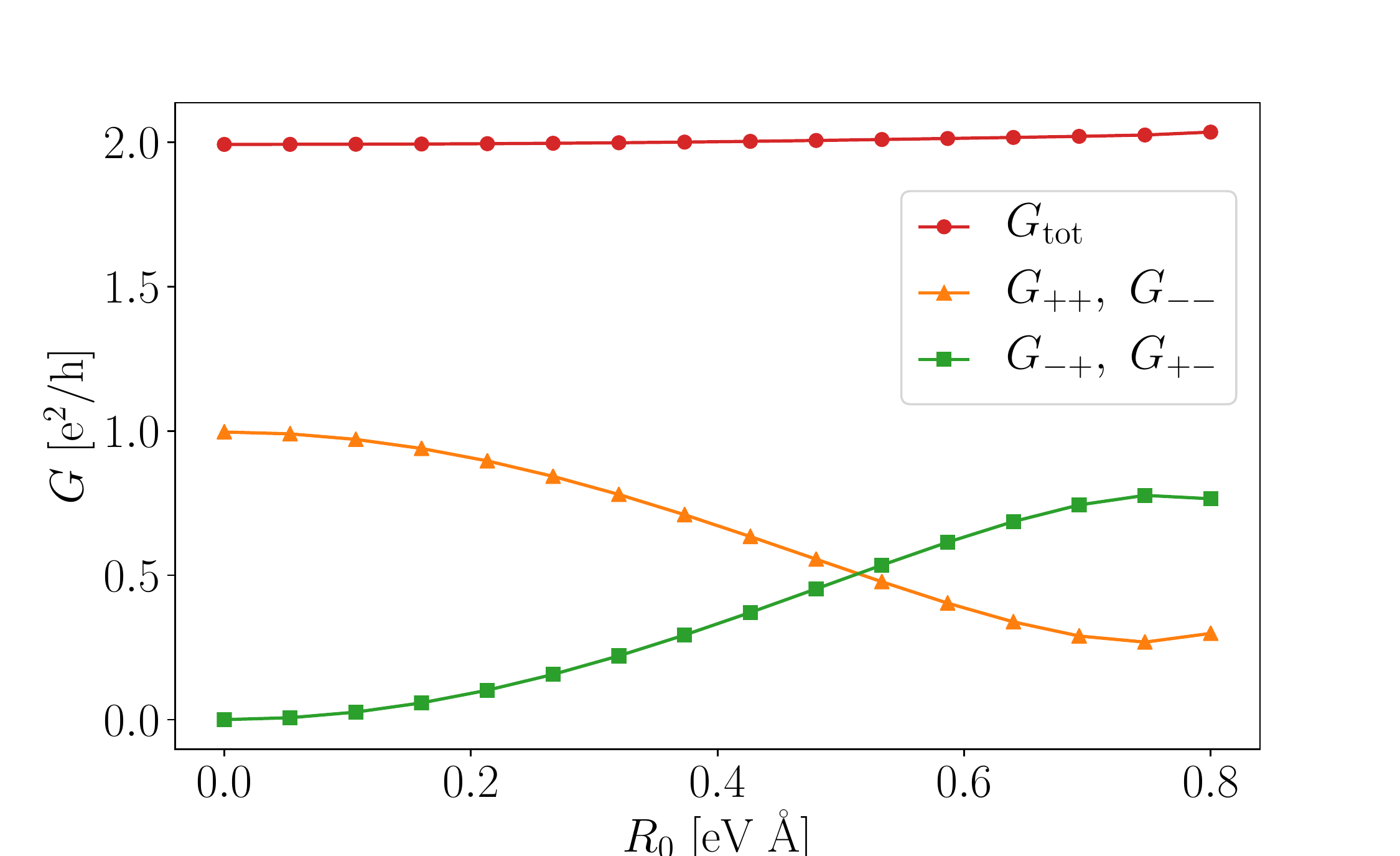}
     \caption{Total conductance $G_{\mathrm{tot}}$ and polarised spin conductances $G_{i j}$ with $i,j=\pm$ as a function of $R_0$. The systems is a slab of \ce{Na_3Bi} with dimensions $150 \times 150 \times 100 ~\si{\angstrom^3}$ and $L_{\mathrm{RSOC}} = 50~\si{\angstrom}$.
     %
     }
    \label{fig:rNa3Bi:CondSwitch}
\end{figure}

In a slab geometry of an ideal DSM, electron transport by surface states shows a quantized conductance of $2 e^2/h$ due to the contribution of the states with opposite chirality in the two surfaces. In fact, scattering states in the absence of the RSOC interaction show two channels, one spin surrounds the cuboid while the other propagates in the plane $y=0$.
After the inclusion of the RSOC, a spin-rotation effect can arise in the $y=0$ plane and hence, due to the mixing induced by RSOC, the spin-flip conductance $G_{\pm \mp}$ becomes nonzero and a spin-switch device is obtained. In figure~\ref{fig:rNa3Bi:CondSwitch} the spin polarized conductance is shown along with the total conductance for a slab of \ce{Na_3Bi}. The RSOC term is implemented using a step function that becomes zero for $y>L_{\mathrm{RSOC}}.$ 
Smoother functions have been tested without significant deviations from the presented results due to the highly peaked shape of the surface states near the interface. 
\revision{In fact, the value of the length  $L_{\mathrm{RSOC}}$ is connected to the charge accumulation near the surface that breaks the symmetry. In our system, the charge accumulation is related to the surface states and their decay. Therefore, the proper length is of the order of the localization length of the surface states. We observe that this assumption yields values of the order of $L_{\mathrm{RSOC}} \sim 50~\si{\angstrom}$.
}

Figure \ref{fig:rNa3Bi:CondSwitch} shows the conductance of the proposed spin switcher. Notice that the total conductance does not decrease and the spin-switch effect becomes dominant with respect to the spin-conserved current for sufficiently high values of $R_0$. 

\begin{figure}[h]
    \centering
    \includegraphics[height=5cm]{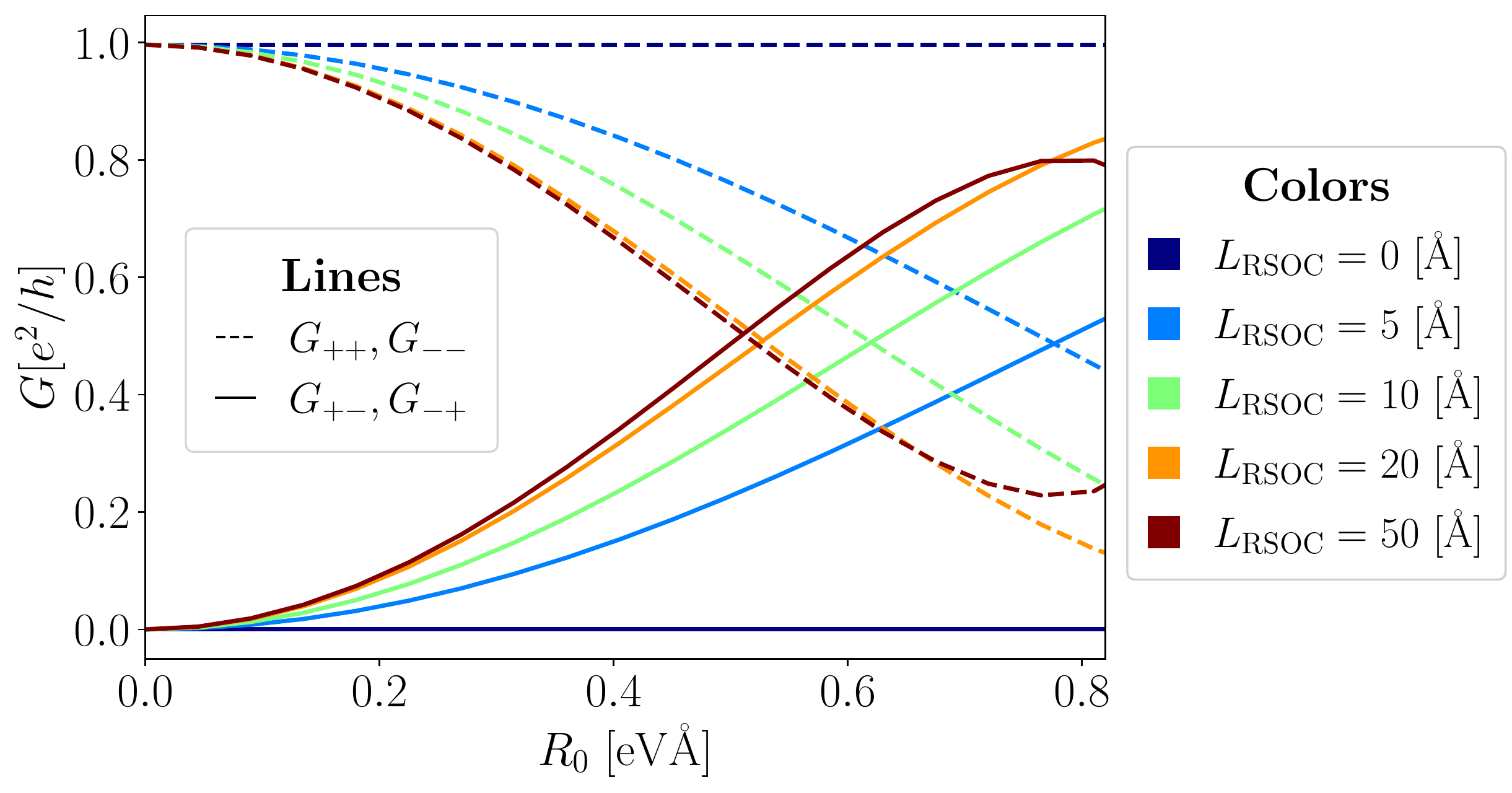}
    \caption{Polarized spin conductance $G_{\pm \pm}$ (in dashed line) and $G_{\pm \mp}$ (in solid line) for a range of $L_\mathrm{RSOC}$ highlighted with the colors defined in the legend. The legend $L_\mathrm{RSOC} = 0~ \si{\angstrom}$ corresponds to the case with no Rashba coupling.}
    \label{fig:rNa3Bi:LRSOC}
\end{figure}

\revision{Notice that an accurate determination of the length $L_\mathrm{RSOC}$ is only possible within DFT calculations that are outside the scope of this work. However, we have checked that the value of $L_{\mathrm{RSOC}}$ is not critical, as long as is larger than the decay length of the surface states. Figure \ref{fig:rNa3Bi:LRSOC} shows the conductance for a range of values of $L_\mathrm{RSOC}$ as a function of $R_0$ showing that the spin-switch effect is present for increasing $R_0$ matching smaller values of the Rashba length. 
}

\subsection{Effect of the quadratic terms and remarks on the system size}

\begin{figure}[htb]
    \centering
    \includegraphics[height=5cm]{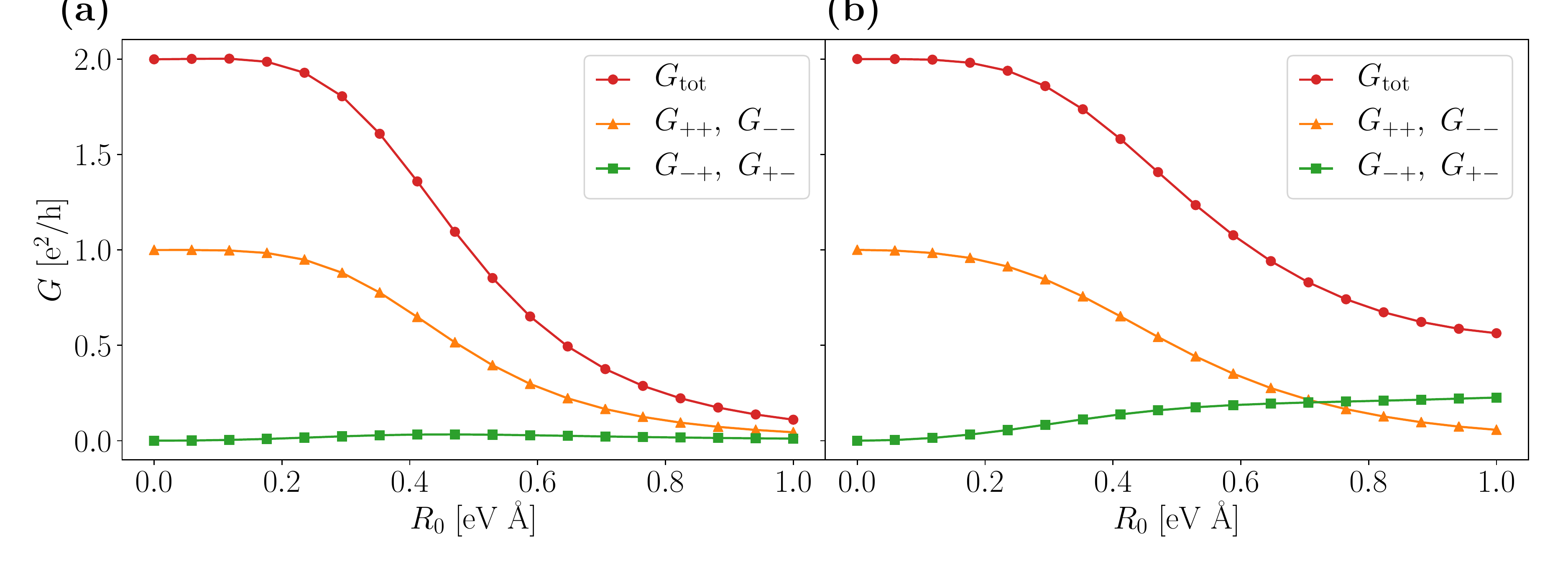}
    \caption{Total conductance $G_{\mathrm{tot}}$ and polarised spin conductance $G_{i j}$ ($i,j= \pm$) as a function of the RSOC coupling constant $R_0$. 
    (a)~Slab of DSM with the same parameters as \ce{Na_3Bi} but setting the quadratic terms to zero, i.e. with $c_0 = c_1 = c_2 = 0$. The dimensions are $150\times 150\times 100 ~\si{\angstrom^3}$.
    (b)~\ce{Na_3Bi} sample of size $80 \times 150 \times 40 ~\si{\angstrom}$ and $L_{\mathrm{RSOC}} = 50~ \si{\angstrom}$.}
    \label{fig:Cond:Others}
\end{figure}

The non-trivial dispersion of the surface states in the $z$-direction has an important role in the design of a spin-switcher. In fact, the absence of quadratic terms in the Hamiltonian leads to a trivial rotation of the spin-chiral basis without spin-flip effects. This can be seen in figure~\ref{fig:Cond:Others}(a), where the conductance in the absence of quadratic terms is reported. In this case, the basis rotates with increasing $R_0$ and the injected current is increasingly non-polarized within the new spin basis. As long as spin-flip is absent, the spin-polarized current as well as the total current decrease with increasing $R_0$, while the spin-flip conductance always vanishes.


Moreover, the magnitude of the effect depends much on the particular design of the device and thinner slabs typically show lower values for the total conductance. This is due to the multiplicity of the surface states of the system. In fact, due to the finite size of the samples and the existence of 2D surface states, the slab shows multiple surface modes that come from the quantization of the momenta as was discussed before and shown in figure~\ref{fig:trans:bulkVsSurf}. In bigger samples, they are closer in energy and  many surface modes can take part in the electron transport, thus leading to higher vales of conductance. In smaller samples, like the one represented in figure~\ref{fig:Cond:Others}(b), the spin-flip effect is also noticeable but the conductance is lowered as long as the next surface mode is separated in energy. A careful adjustment of the Fermi energy for each value of $R_0$ would prevent this conductance reduction. However, we have chosen to leave the Fermi energy fixed as this would provide a simpler setting for actual experiments.

\section{Effect of impurity disorder} \label{sec:Disorder}

\begin{figure}
    \centering
    \includegraphics[height=5.5cm]{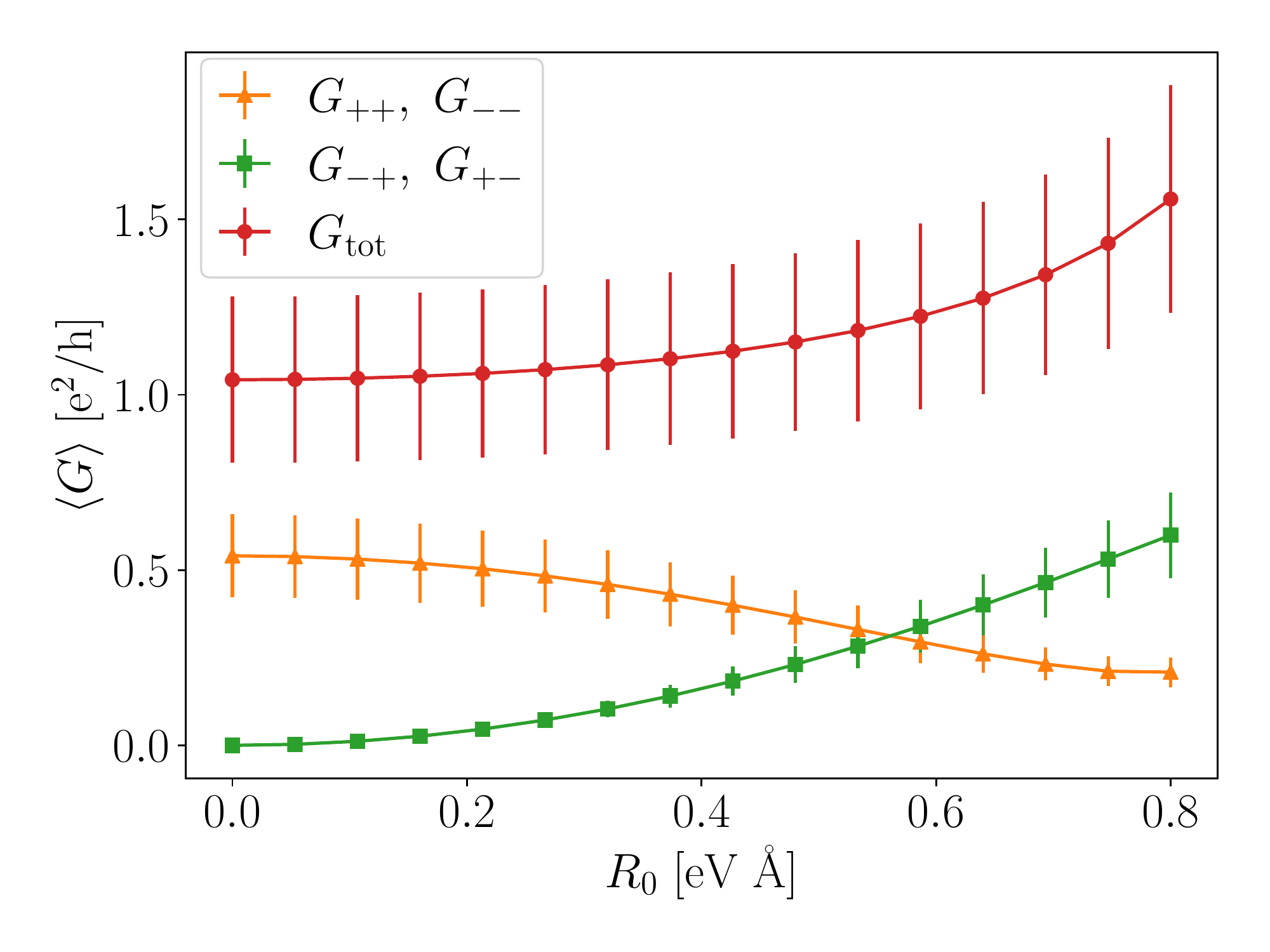}
    \caption{Average conductance in the presence of point-like impurities of strength $W_0=0.5\,\si{eV}$ and density $n_\mathrm{imp} = 4 \times 10^{19} ~\si{cm^{-3}}$. The rest of parameter are the same of figure~\ref{fig:rNa3Bi:CondSwitch}. Results are averaged over $250$ realizations of disorder and error bars correspond to the standard deviation. }
    \label{fig:Na3Bi:Disorder}
\end{figure}

\revision{
In order to check the robustness of the spin-switch effect against disorder in a more realistic scenario, we have performed transport simulations with point defects placed randomly in the scattering region of the setup. We chose non-magnetic point-like impurities with an equal weight in the four orbitals of the basis of the Hamiltonian~\eqref{eq:MM:HamD}. The simulations proceed as follows. We take at random a fraction of sites of the grid used in the numerical solution of the transport problem. An amount $W_0$ is added to the site energy of these random sites. The strength of the impurity potential $W_0$ is taken to be comparable to the intra-band hopping in order to have an impact on the electron states. Specifically, we set $W_0 = 0.5~\si{eV}$ in our simulations and calculate the spin polarised conductance by averaging over $250$ realizations of disorder.
Figure \ref{fig:Na3Bi:Disorder} shows the average conductance for the same parameters of figure~\ref{fig:rNa3Bi:CondSwitch} when the impurity density is as large as $n_\mathrm{imp} = 4 \times 10^{19} ~\si{cm^{-3}}$. Error bars indicate the standard deviation. We see that the spin-flip effect is still clearly revealed. The only noticeable effects are an overall reduction of the conductivity, as expected, and a slight shift of the crossing point towards higher values of $R_0$. Therefore, we can confidently assert that the effect is quite robust against disorder, even at such large impurity density. 
}

\section{Conclusions and prospective research} \label{sec:Concl}


We have studied the effect of RSOC in the behavior of the surface Fermi arcs of Dirac semimetals. We have considered two models for the description of the Dirac semimetals, namely a minimal model which describes two isotropic Dirac cones with totally flat Fermi arcs and a realistic models that has extra dispersive terms, describing the low energy physics of materials like Na$_3$Bi and Cd$_3$As$_2$ \cite{Crasse2018}. By studying the effective Hamiltonian for the surface states of a semiinfinite block of the material, we have obtained analytical results concerning the mixing of the Fermi arcs of different chiralities in the same surface. The surface states of Dirac semimetals when RSOC terms are not included have definite chirality, its spin orientation is constant and become energy independent. Once RSOC is taken into account, the spinors of counter-propagating surface states are still orthogonal but acquire some non-trivial spin structure. We have also calculated the modification of the velocity due to the RSOC. 

The new spinors can be parametrized by a momentum-dependent rotation. This spin rotation is reflected in the properties of the spin-dependent conductance in a two-terminal setup. We have studied a bar of the material connected to two-dimensional metallic leads in one of the surfaces. Due to the electron transport through the topological Fermi arcs, the conductance without RSOC is quantized, with equal contribution from both spins and without signatures of spin-flip conductance. However, when the RSOC is switched on, the quantization is lost and the total conductance varies. The spin-flip part of the conductance can be greater than the spin-conserved conductance at larger values of the coupling parameter $R_0$, due to the spin-rotation effects that arise from the RSOC. For higher values of $R_0$, the injected polarized current mainly rotates, thus inverting the spin polarization of the current. This is found in Na$_3$Bi while for the minimal model the spin-flip is absent due to a trivial rotation of the spin basis. Therefore, this system could be used as a spin switcher for a suitable set of parameters. A setup where the value of the RSOC and the spin switching effect can be controlled by an external gate is very promising. 

Notice that the effect of the Rashba term is completely different for the minimal model without quadratic terms and the realistic DFT-fitted model for \ce{Na_3Bi}. In fact, the inclusion of the quadratic terms turns out to be crucial. In simple theoretical approaches the quadratic terms are usually neglected. However, they are essential for the correct understanding of the spin-switcher device proposed in this work. 

\revision{The effect of the disorder is also quantitatively addressed in order to prove the robustness of the spin-switcher. We find that this effect is resilient to relatively high density of point-like disorder as shown in section \ref{sec:Disorder}. In fact, the increasing of the impurity density, decreases the total conductance and displaces the crossing point in $R_0$ of the opposite polarized conductance without destroying the effect for impurity densities of the order of $n_\mathrm{imp} \sim 10^{19} ~\si{cm^{-3}}$}.

Inelastic scattering effects is outside the scope of our work. However, the breaking of the axial spin symmetry might lead to an increased inelastic backscattering that can modify the electrical response of the material. The analytical results were obtained in a semi-infinite system. Understanding the importance of the coupling between surface states in realistic finite-size systems would be important for the design of actual devices and will be addressed in future works. Also, surface reconstruction and relaxation effects due to the growth of the topological material on the substrate should be taken into account but would obviously depend on the particular experimental realization. Moreover, the study of the system in a four terminal Hall configuration would be an interesting complementary work in order to elucidate the role of the transverse current.

In conclusion, we believe that the presented results are solid and pave the way to future research. Specially, they allow us to foresee possible applications of the RSOC interaction in the design of spintronic devices based on topological semimetals. 

\ack

We acknowledge financial support through Spanish grants PGC2018-094180-B-I00 (MCIU/AEI/FEDER, EU), MAT2016-75955, MAT2017-86717-P and PID2019-106820RB-C21 (MINECO/FEDER, EU), CAM/FEDER  Project  No. S2018/TCS-4342 (QUITEMAD-CM) and CSIC Research Platform PTI-001.  
G. P.  also acknowledges support from the DFG within SBF 1277 and from the “UA: Matem\'{a}tica Aplicada a la Teor\'{i}a de la Materia Condensada” with the Carlos III University.
\section*{References}

\bibliography{references}
\end{document}